\documentclass[12pt]{article}
\usepackage{epsfig}
\usepackage{epsf}
\usepackage{bm}
\begin{document}

\vspace{2cm}

{\large \sf

~~ \vspace{2cm}

\begin{flushright}
cond-mat/0602009
\end{flushright}
}

\begin{center}

{\huge \sf
Comments on the Superconductivity Solution  \\
of an Ideal Charged Boson System$^*$}

\vspace{3cm}

{\large \sf

R. Friedberg$^1$ and T. D. Lee$^{1,~2}$\\
} \vspace{.5cm}

{\sf
{\it 1. Physics Department, Columbia University}\\
{\it New York, NY 10027, U.S.A.}\\
{\it 2. China Center of Advanced Science and Technology (CCAST/World Lab.)}\\
{\it P.O. Box 8730, Beijing 100080, China}\\
}
\end{center}

\vspace{1cm}

\begin{center}

{\large \sf Abstract}
\end{center}

\vspace{.5cm}

{\large \sf ~~~~ We review the present status of the
superconductivity solution

~~ for an ideal charged boson system, with suggestions for
possible

~~ improvement.}

\vspace{3cm}

----------------------------------
{\sf

A dedication in celebration of the 90th birthday of Professor V.
L. Ginzburg

\noindent * ~This research was supported in part by the U. S.
Department of Energy Grant

 DE-FG02-92ER-40699 }

\newpage

\begin{center}
 {\large \bf 1. Introduction}
\end{center}

{\large \sf

An ideal charged boson system is of interest because of the
simplicity in its formulation and yet the complexity of its
manifestations. The astonishingly complicated behavior of this
idealized system may provide some insight to the still not fully
understood properties of high $T_c$ superconductivity. As is well
known, R. Schafroth[1] first studied the superconductivity of this
model fifty years ago. In this classic paper he concluded that at
zero temperature $T=0$ and in an external constant magnetic field
$H$, there is a critical field
$$
(H_c)_{Sch}=e\rho/2m\eqno(1.1)
$$
with $\rho$ denoting the overall number density of the charged
bosons and $m$, $e$ their mass and electric charge respectively;
the system is in the super phase when $H<H_c$, and in the normal
phase when $H>H_c$. Due to an oversight, Schafroth neglected the
exchange part of the electrostatic energy, which invalidates his
conclusion as was pointed out in a 1990 paper [2] by Friedberg,
Lee and Ren (FLR). This oversight when corrected makes the ideal
charged boson model even more interesting. Some aspects of this
simple model are still not well understood.

In what follows we first review the Schafroth solution and then
the FLR corrections. Our discussions are confined only to $T=0$.

%\newpage

\begin{center}
 {\large \bf 2. Hamiltonian and Schafroth Solution}
\end{center}

Let $\phi({\bf r})$ be the charged boson field operator and
$\phi^\dag({\bf r})$ its hermitian conjugate, with their
equal-time commutator given by
$$
[\phi({\bf r}),~\phi^\dag({\bf r}')]=\delta^3({\bf r}-{\bf
r}').\eqno(2.1)
$$
These bosons are non-relativistic, enclosed in a large cubic
volume $\Omega=L^3$ and with an external constant background
charge density $-e\rho_{ext}~$ so that the integral of the  total
charge density
$$
eJ_0\equiv e \phi^\dag\phi-e \rho_{ext}\eqno(2.2)
$$
is zero. The Coulomb energy operator is given by
$$
H_{Coul}=\frac{e^2}{8\pi} \int~|~{\bf r}-{\bf r}'|^{-1}~:J_0({\bf
r})J_0({\bf r}'):~ d^3rd^3r'\eqno(2.3)
$$
where $:~:$ denotes the normal product in Wick's notation[3] so as
to exclude the Coulomb self-energy.

Expand the field operator $\phi({\bf r})$ in terms of a complete
orthonormal set of $c$-number function $\{f_i({\bf r})\}$:
$$
\phi({\bf r})=\sum\limits_i a_if_i({\bf r})\eqno(2.4)
$$
with $a_i$ and its hermitian conjugate $a_i^\dag$ obeying the
commutation relation $[a_i,~a_j^\dag]=\delta_{ij}$, in accordance
with (2.1). Take a normalized state vector $|>$ which is also an
eigenstate of all $a_i^\dag a_i$ with
$$
a_i^\dag a_i |>=n_i|>.\eqno(2.5)
$$
For such a state, the expectation value of the Coulomb energy
$E_{Coul}$ can be written as a sum of three terms:
$$
<|H_{Coul}|>=E_{ex}+E_{dir}+E_{dir}'\eqno(2.6)
$$
where
$$
E_{ex}=\sum\limits_{i\neq j}\frac{e^2}{8\pi}\int d^3rd^3r' |~{\bf
r}-{\bf r}'|^{-1}n_i n_j f_i^*({\bf r}) f_j^*({\bf r}')f_i({\bf
r}')f_j({\bf r})
$$
$$
E_{dir}=\frac{e^2}{8\pi}\int d^3rd^3r' |~{\bf r}-{\bf
r}'|^{-1}<|J_0({\bf r})|><|J_0({\bf r}')|>\eqno(2.7)
$$
and
$$
E_{dir}'=-\sum\limits_{i}\frac{e^2}{8\pi}\int d^3rd^3r'|~{\bf
r}-{\bf r}'|^{-1}n_i |f_i({\bf r})|^2 |f_i({\bf r}')|^2.
$$
The last term $E_{dir}'$ is the subtraction, recognizing that in
Wick's normal product each particle does not interact with itself.

In the Schafroth solution, for the super phase at $T=0$ all
particles are in the zero momentum state; therefore, on account of
(2.2) the ensemble average of $J_0$ is zero and so is the Coulomb
energy. For the normal phase, take the magnetic field ${\bf
B}=B\hat{z}$ with $B$ uniform and pick its gauge field ${\bf
A}=Bx\hat{y}$. At $T=0$, let
$$
f_i({\bf r})=e^{ip_iy}\psi_i(x).\eqno(2.8)
$$
Schafroth assumed $p_i=eBx_i$ with $x_i$ spaced at regular
intervals $\lambda=2\pi/eBL$, which approaches zero as
$L\rightarrow \infty$. This makes the boson density uniform and
therefore $E_{dir}=0$. In the same infinite volume limit, one can
show readily that $\Omega^{-1}E_{dir}'\rightarrow 0$. Since
Schafroth omitted $E_{ex}$, his energy consists only of
$$
E_{field}=\int d^3r \frac{1}{2}~B^2,\eqno(2.9)
$$
$$
E_{mech}=\sum\limits_{i}n_i \int d^3r
\frac{1}{2m}~\bigg(\frac{d\psi_i}{dx}\bigg)^2\eqno(2.10)
$$
and
$$
E_{dia}=\sum\limits_{i}n_i \int d^3r \frac{1}{2m}~(p_i-eA_y(x))^2
$$
$$
~~~~~~~~=\sum\limits_{i}n_i \int d^3r
\frac{eB}{2m}~(x-x_i)^2.\eqno(2.11)
$$
The sum of (2.10) and (2.11) gives the usual cyclotron energy
$$
E_{mech}+E_{dia}=\sum\limits_{i}n_i ~\frac{eB}{2m}.\eqno(2.12)
$$
Combining with (2.9), Schafroth derived the total Helmholtz free
energy density in the normal phase at zero temperature to be
$$
F_n=\frac{1}{2}~B^2+\frac{e\rho}{2m}~B\eqno(2.13)
$$
(Throughout the paper, we take $e$ and $B$ to be positive, since
all energies are even in these parameters.)

The derivation of (2.13) is, however, flawed by the omission of
$E_{ex}$. It turns out that for the above particle wave function
(2.8), when $x_i-x_j$ is $<<$ the cyclotron radius
$a=(eB)^{-\frac{1}{2}}$, the coefficient of $n_in_j$ in $E_{ex}$
is proportional to $|x_i-x_j|^{-1}$. Hence $\Omega^{-1}E_{ex}$
becomes $\infty$ logarithmically as the spacing $\lambda
\rightarrow 0$.

\begin{center}
 {\large \bf 3. Corrected Normal State at High Density}
\end{center}

In this and the next section, we review the FLR analysis for the
high density case, when $\rho > r_b^{-3}$ where $r_b=$ Bohr radius
$=4\pi/me^2$.

\vspace{1cm}

\noindent a. \underline{Strong field}. We discuss first the case
when $B$ is $>>(m\rho)^{\frac{1}{2}}$, so that the Coulomb
correction to the magnetic energy (2.13) can be treated as a
perturbation. To find the groundstate energy, we shall continue to
assume (2.8) with $p_i=eBx_i$ and $x_i$ equally spaced at interval
$\lambda$, but keeping $\lambda \neq 0$. Now as $\Omega
\rightarrow \infty$, $\Omega^{-1}E_{dir}'$ remains zero, but
$\Omega^{-1}E_{dir}$ in fact increases as $\lambda^2$ for
$\lambda>>a$, the cyclotron radius. The lowest value of
$E_{dir}+E_{ex}$ are both complicated in this range. The
minimization can be done exactly, yielding
$$
\lambda=\pi a \lambda_0\eqno(3.1)
$$
where
$$
1-\bigg(\frac{\pi}{2}\bigg)^{\frac{1}{2}} \lambda_0 = \lambda_0^2
\sum\limits_{\mu=1}^{\infty}e^{-2\mu^2/\lambda_0}\eqno(3.2)
$$
The sum of $E_{dir}$ and $E_{ex}$ is found to be proportional to
$1/B$. Hence, (2.13) is replaced by
$$
F_n=\frac{1}{2}~B^2+\frac{e\rho}{2m}~B+\frac{e\rho^2}{B}~\gamma_n\eqno(3.3)
$$
with
$$
\gamma_n=\gamma_{dir}+\gamma_{ex}=0.00567+0.00715=0.0128.\eqno(3.4)
$$
\vspace{.5cm}

\noindent b. \underline{Weak field}. Clearly (3.3) cannot be
extended to $B\rightarrow 0$, as the last term would diverge. In
its derivation the $\psi_i(x)$ in (2.8) is taken to be the usual
simple harmonic oscillator wave function determined by the
magnetic field $B$ only, without regard to $E_{Coul}$. This is
valid only when $B>>(m\rho )^{\frac{1}{2}}$. For much smaller $B$,
we may consider a configuration in which the function $\psi_i(x)$
is spread out flat over a width $\lambda-l$, then drops to zero
sinusoidally over a smaller width $l$ on each side. Neighboring
$\psi_i$ overlap only in the strips of width $l$.Thus, it can be
arranged that $\sum\limits_i |\psi_i(x)|^2$ is uniform and hence
$$
E_{dir}=0.\eqno(3.5)
$$

%\newpage

For sufficiently weak $B$, we find $\lambda>>$ the London length
$\lambda_L=(e^2\rho/m)^{-\frac{1}{2}}$. We must then drop the
assumption that $B$ is uniform; it is largest in the overlap
region and drops to zero over the length $\lambda_L$ in either
side. Let $\overline{B}$ be the average of $B$ over $\Omega$ and
$\overline{a} \equiv (e\overline{B})^{-\frac{1}{2}}$ the
corresponding "cyclotron radius". It is then found that
$$
\lambda =\frac{\pi \overline{a}^2}{\epsilon \lambda_L},\eqno(3.6)
$$
$$
l=\lambda_L \epsilon^2 <<\lambda_L\eqno(3.7)
$$
where
$$
\epsilon=\bigg(\frac{\pi^5}{8m^2\lambda_L^2}\bigg)^{1/7}<<1
\eqno(3.8)
$$
is independent of $\overline{B}$. The energies $E_{mech}$,
$E_{ex}$, $E_{dia}$ are all proportional to $\overline{B}$, and
one obtains for the free energy density
$$
F_n=\frac{1}{2}~\overline{B}^2+\frac{e\rho}{m}~\eta
\overline{B}\eqno(3.9)
$$
with
$$
\eta=\frac{7\pi}{16\epsilon}>>1,\eqno(3.10)
$$
much bigger than the Schafroth result.

\noindent c.\underline{Intermediate field}. Let
$$
B_1=e\rho/m,~~~B_2=(m\rho)^{\frac{1}{2}}\eqno(3.11)
$$
Between the above strong field case $B>>B_2$ and the weak field
case when $\overline{B}<<\eta B_1$, we have the regime when $\psi$
remains flat as in the previous section, but with $\lambda_L >>
\lambda \geq l$. Hence, $E_{dir}=0$ as in (b), but $B$ is uniform
as in (a). One obtains an estimate
$$
F_n=\frac{1}{2}~B^2+\frac{5}{32}~B_0^{7/5}B^{3/5}\eqno(3.12)
$$
where
$$
B_0=\bigg(\frac{16}{3}\pi m \lambda_L\bigg)^{2/7}B_1.\eqno(3.13)
$$

The formulas (3.3-4), (3.9-10) and (3.12-13) are strictly upper
bounds, which might be improved with better wave functions. We
hope these to be good estimates.

\newpage

\begin{center}
 {\large \bf 4. Super State at High Density}
\end{center}

\noindent a. \underline{$H_{c_1}$}. The coherent length $\xi$
governing the disappearance of the normal phase outside a vortex
is found from the Ginzburg-Landau (G.-L.) equation[4] to be
$$
\xi=(2\lambda_L/m)^{\frac{1}{2}}.\eqno(4.1)
$$
and hence (taking $m\lambda_L>>1$) $\lambda_L>>\xi$ so that we
should have a type {\rm II} superconductor, with the critical
field for vortex penetration
$$
H_{c_1} \cong \frac{1}{2}B_1[~\ln
(\lambda_L/\xi)+1.623~]\eqno(4.2)
$$
in which the constant differs from that given by the G.-L.
equation because of the long range Coulomb field.

However, in Schafroth's solution because of (2.13), his normal
phase would begin to exist at
$$
(H_c)_{Sch}=\frac{1}{2}~B_1<<H_{c_1},\eqno(4.3)
$$
above which $(F_n-BH)_{Sch}$ would also be lower than that for the
super phase, making it a Type {\rm I} superconductor. But
Schafroth's solution is invalid; (2.13) must be replaced by (3.9)
at low $B$, giving
$$
H_c=\eta B_1\eqno(4.4)
$$
with $\eta>>\frac{1}{2}$, as shown in (3.10).

Next, we compare the above corrected $H_c$ with $H_{c_1}$. Using
(4.1), we write (4.2) as
$$
H_{c_1}=\bigg(\frac{7}{8}~\ln\zeta\bigg)B_1\eqno(4.5)
$$
where
$$
\ln \zeta =\frac{4}{7}~\bigg(\frac{1}{2}\ln
\frac{m\lambda_L}{2}+1.623\bigg).\eqno(4.6)
$$
Likewise, because of (3.8) and (3.10), the parameter $\eta$ in
(4.4) can also be expressed in terms of the same $\zeta$:
$$
\eta=\kappa~\zeta\eqno(4.7)
$$
with the constant $\kappa$ given by
$$
\kappa=\frac{7}{8}~\bigg(\frac{\pi}{2}~e^{-3.246}\bigg)^{2/7}.\eqno(4.8)
$$
Thus,
$$
\frac{H_c}{H_{c_1}}=\frac{8\kappa}{7}~\frac{\zeta}{\ln
\zeta}.\eqno(4.9)
$$
Now $\zeta/\ln \zeta$ has a minimum$=e$ when $\zeta=e$. Thus,
$$
\frac{H_c}{H_{c_1}}>\bigg(\frac{\pi}{2}\bigg)^{2/7}e^{0.07}>1,\eqno(4.10)
$$
and the system is indeed a Type {\rm II} superconductor.

\noindent b. \underline{Vortices}. Once $H=H_{c_1}+$, the vortices
appear and soon become so numerous that their typical separation
is of the order of $\lambda_L$. This gives an average $B$ of the
order of $B_1=e\rho/m$.

\noindent c. \underline{$B>B_1$}. To increase $B$ further, it is
necessary to increase $H$ on account of the interaction energy
between vortices. The vortex separation distance is of the order
of the cyclotron radius $a=(eB)^{-\frac{1}{2}}$. In the regime
$\xi<<a<<\lambda_L$ (correspondingly, $B_2>>B>>B_1$), the vortices
naturally form a lattice to minimize their interaction energy. An
involved calculation gives the Helmholtz free energy density at
$T=0$ to be
$$
F_s=\frac{1}{2}B^2+\frac{1}{4}B_1B(\ln \frac{B_2}{B}+{\sf
const})\eqno(4.11)
$$
where $B_2=(m\rho)^{\frac{1}{2}}$ and the constant is
\begin{eqnarray*}\label{4.12}
~~~~~~~~~~-\ln 4\pi+\left\{
\begin{array}{ll}
4.068~,~~~{\sf square~~lattice}\\
4.048~,~~~{\sf triangular~~lattice}.
\end{array}
\right.~~~~~~~~~~~~~~~~~~~~~~~~~~(4.12)
\end{eqnarray*}

\noindent d. \underline{$B>>B_2$}. In this case of very strong
magnetic field when the cyclotron radius and the separation
distance are both much less than $\xi$ (but we assume that the
system remains non-relativistic). The free energy density is
dominated by the RHS of (2.13). However, there is still a super
phase whose wave function is assumed to be given by the Abrikosov
solution, giving by Eq.(8) of Ref.[5] and its Coulomb energy is
calculated as a perturbation. The result for the Helmholtz energy
density in the super phase is
$$
F_s=\frac{1}{2}B^2+\frac{e\rho}{2m}B+\frac{e\rho^2}{B}\gamma_s\eqno(4.13)
$$
with
\begin{eqnarray*}\label{4.14}
~~~~~~~~~~~~\gamma_s=\left\{
\begin{array}{ll}
0.01405~,~~~{\sf square~~lattice}\\
0.01099~,~~~{\sf triangular~~lattice}.
\end{array}
\right.~~~~~~~~~~~~~~~~~~~~~~~~~~~~(4.14)
\end{eqnarray*}

\noindent e. \underline{$H_{c_2}$}. The super phase regimes of
Sections 4a-b, 4c, 4d correspond  (with respect to the value of
$B$ or its average $\overline{B}$) to those of the normal phase
regimes discussed in 3b, 3c and 3a respectively. Since, for the
triangular lattice, a comparison of (4.14) with (3.4) gives
$\gamma_s=0.01099<\gamma_n=0.0128$, we see that $F_s<F_n$ for the
same $B>>B_2$. Similarly, $F_s<F_n$ for the same $B$ in the
regimes $B_2>>B>>B_1$ and $B<<B_1$. From these results and that
$H=dF/dB$ is monotonic in $B$, one can readily deduce that the
Legendre transform
$$
\tilde{F}=F-BH\eqno(4.15)
$$
satisfies $\tilde{F}_s<\tilde{F}_n$ for the same $H$ in all these
regions. (See Figure 1)

From this it seems possible that
$$
H_{c_2}=\infty;\eqno(4.16)
$$
the super phase may persist at high density for all values of the
magnetic field.

\noindent f. \underline{Remarks}. In the problem discussed in
Ref.[5], the Ginzburg-Landau function $\Psi$ is an order
parameter, whereas our $f_i(r)$ are single particle wave
functions. Nevertheless, except for the constant in (4.2), the two
problems have the same physics content at high $\rho$ when
$H<<B_2$. For higher field when $H\geq B_2$, the Ginzburg-Landau
$\Psi$ should vanish; however, this is not true in our problem. At
$T=0$, we place all the particles in the coherent state, making
the charge density to vary greatly within a unit lattice cell. Our
result (4.14) favoring a triangular lattice is unrelated to that
of [5], because the ratio parameter $<|\Psi|^4>/<|\Psi|^2>^2$ in
[5] does not appear in our problem. The lattice dependence in our
problem is electrostatic in origin.

\newpage

\begin{center}
 {\large \bf 5. Low Density at Zero Field}
\end{center}

\noindent a. \underline{Normal Phase}. At very low density and
with zero magnetic field, $E_{dir}'$ of (2.7) becomes important.
The lowest energy is now achieved by placing the individual
charges in separate cells forming a lattice, with little or no
overlap. Hence $E_{ex}$ can be disregarded, and a trial wave
function leads in the limit $\rho\rightarrow 0$ to
$$
N^{-1}(E_{Coul}+E_{mech})=-\alpha_n\frac{e^2}{4\pi R}\eqno(5.1)
$$
where
$$
(4\pi/3) R^3=\rho^{-1}~~~{\sf and}~~~\alpha_n\cong 0.9\eqno(5.2)
$$
very closely. The above formula (5.1) is valid for
$\rho<<r_b^{-3}$, with $r_b$ the Bohr radius.

\noindent b. \underline{Super Phase}. In the same limit, the super
phase energy also becomes negative, as shown by a Bogolubov-type
transformation[6-8]. This leads to
$$
N^{-1}(E_{Coul}+E_{mech})=-\alpha_s\frac{e^2}{4\pi R}\eqno(5.3)
$$
with
$$
0.316<\alpha_s<0.558.\eqno(5.4)
$$
Thus, $\alpha_s<\alpha_n$ and the normal phase holds at
$\rho<<r^{-3}_b$.

\noindent c. \underline{Critical Density}. As $\rho$ increases,
(5.1-2) serves only as a lower bound; i.e.,
$$
N^{-1}(E_{Coul}+E_{mech})>-(0.9)\frac{e^2}{4\pi R}.\eqno(5.5)
$$
For the normal phase, when $\rho$ approaches $r^{-3}_b$, the
single particle wave function leading to (5.1-2) can no longer fit
without overlap. We confine each particle within a cube, give it a
$r^{-1}\sin qr$ wave function as a trial function, just avoiding
overlap so that $E_{ex}=0$. With approximation neglecting the
distinction between sphere and cube, we find
$$
N^{-1}E_{mech}=\frac{\pi^2}{2mR^2},~~~N^{-1}E_{Coul}=-\frac{e^2}{4\pi
R}K_n\eqno(5.6)
$$
where
$$
K_n\approx 0.76~.\eqno(5.7)
$$
Equating the above $N^{-1}(E_{Coul}+E_{mech})$ for the normal
phase with the corresponding expression (5.3) for the super phase,
we find the critical density $\rho_c$ given by
$$
r_b^{3}~\rho_c=\frac{6}{\pi^7}~(K_n-\alpha_s)^3.\eqno(5.8)
$$
The system is in the normal state when $\rho<\rho_c$, and in the
super state when $\rho>\rho_c$. (Eq.(4.12) in  the FLR paper is
equivalent to (5.8), but without the subtraction of $K_n$ by
$\alpha_s$.)

\begin{center}
 {\large \bf 6. Further Improvement}
\end{center}

Although the FLR paper (66 pages in the Annals of Phys.) is quite
lengthy, several important questions remain open.

\noindent a. The energies in above sections 3 and 4 are all upper
bounds obtained from trial functions. Perhaps a better trial
function, like changing slabs into cylinders, might lower these
bounds and put into questions some of the FLR conclusions. Also a
numerical calculation exploring the transition regions would be
valuable in case there are surprises, particularly when $B\sim
B_2$. In this connection we note that, e.g., in (4.9) the relevant
factor in $\zeta/\ln \zeta$ is $(m\lambda_L)^{2/7}/\ln
(m\lambda_L)$, which becomes large when $m\lambda_L \rightarrow
\infty$; yet, it is $<1$ when $m\lambda_L=100$ and only near but
still less than $2$ when $m\lambda_L$ is $2000$.

\noindent b. The calculation of the above (5.6-7), i.e., Section
4.2 in the FLR paper, can be improved in several ways. First,
consider the integral $\frac{1}{2}\int J_0Vd^3r$, with $V$ the
potential due to $J_0$. Because the spatial integral of $J_0$ is
zero, and since each particle does not interact with itself we
have
$$
N^{-1}(E_{dir}+E_{dir}')=e\int\psi^2_0({\bf r})\overline{V}({\bf
r})d^3r\eqno(6.1)
$$
where $\psi_0({\bf r})$ is located inside a sphere, centered at
zero, and $\overline{V}({\bf r})$ is due to all of $J_0$ except
the term due to $\psi^2_0$. Second, there is no need to ignore the
distinction between sphere and cube. Using theorems from
electrostatics, one can reduce (6.1) to the solution of a Madelung
problem with like charges at lattice points and a background
charge filling \underline{all} space, plus a correction
$\frac{1}{6}e^2\rho\int\psi^2_0({\bf r})r^2d^3r$. This correction
can be combined with $E_{mech}$ to optimize $\psi_0$, and the
Madelung problem can be done by known methods. Third, the energy
can probably be reduced by placing the centers of the particle
wave functions on a body-centered cubic lattice, as the cell
available to each particle would then be more nearly spherical
than a cube.

\noindent c. Both FLR and the present paper have left open the
question of what happens at low density and high field. It would
be surprising if the boundary between super and normal phases were
independent of the magnetic field strength. In the $H$ versus
$\rho$ phase diagram at $T=0$ and high $H$, does the boundary
between normal and super phases bend towards lower $\rho$, or
towards higher?

\begin{center}
{\large \bf 7. Comment}
\end{center}

The two most striking results in our paper are $H_{c_1}<H_c$,
making the superconductor Type {\rm II} instead of Type {\rm I},
and that $H_{c_2}$ might be infinite. An improvement in the weak
field normal trial function (the above Section 3b) might
invalidate the first conclusion by lowering $\eta$ in (3.9). An
improvement in the strong field normal trial function (Section 3a)
could invalidate the second conclusion by lowering $\gamma_n$ in
(3.3).

The field of condensed matter physics has received from its very
beginning many deep and beautiful contributions from Russian
physicists and masters L. D. Landau, V. L. Ginzburg, N. N.
Bogolubov, A. A. Abrikosov, A. M. Polyakov and others. It is our
privilege to add this small piece to honor this great and strong
tradition and to celebrate the 90th birthday of V. L. Ginzburg.

%\newpage
\begin{figure}[htb]
\begin{center}%\vspace*{2.0cm}
{\epsfysize=11cm \epsffile{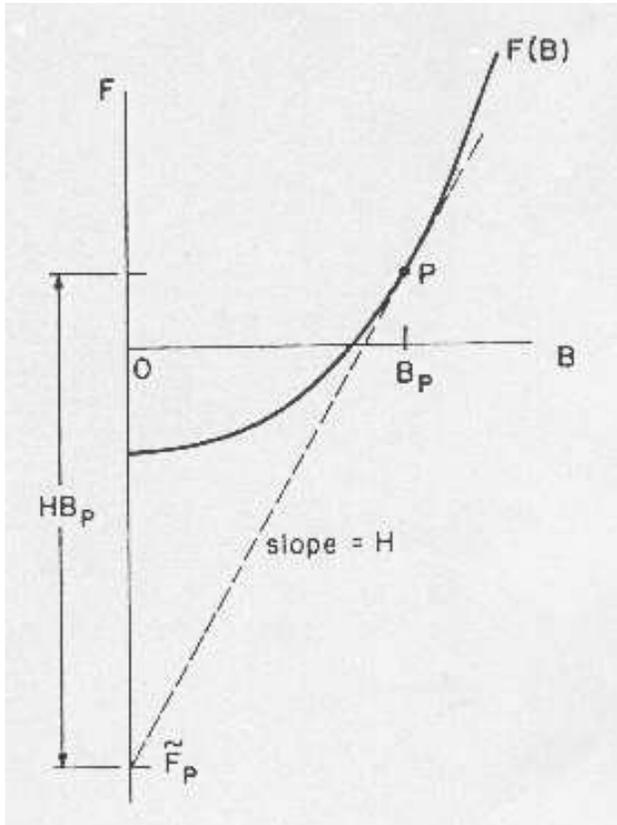}}%
\vglue 0cm\caption{\label{Figure 1}Graphical construction of
$\tilde{F}=F-HB$. At any point $P$ on the curve $F(B)$, the
intercept of its tangent with ordinate gives $\tilde{F}$, since
the tangent has a slope $H$. The subscript $P$ denotes the values
of $B$ and $\tilde{F}$ at $P$.}
\end{center}
\end{figure}

%\newpage

%\begin{center}
%{\large \bf References}
%\end{center}
{\normalsize {

}}

}
\end{document}